\documentclass[conference]{IEEEtran}
\IEEEoverridecommandlockouts
\usepackage{cite}
\usepackage[hyphens]{url}
\usepackage{hyperref}
\usepackage{orcidlink}
\usepackage{comment}
\usepackage{amsmath,amssymb,amsfonts}
\usepackage{algorithm}
\usepackage{algorithmic}
\usepackage{graphicx}
\usepackage{textcomp}
\usepackage{xcolor}
\usepackage{soul}
\usepackage{acronym}
\usepackage{multicol}
\usepackage{multirow}
\usepackage{subfigure}
\usepackage{booktabs}

\usepackage[a4paper, margin=19mm]{geometry}
\usepackage{cite}
\usepackage{textcomp}
\usepackage{amsfonts}

\def\BibTeX{{\rm B\kern-.05em{\sc i\kern-.025em b}\kern-.08em
    T\kern-.1667em\lower.7ex\hbox{E}\kern-.125emX}}

\begin{document}

\title{Balanced Space- and Time-based Duty-cycle Scheduling for Light-based IoT\vspace{-0.1em}}

\author{\IEEEauthorblockN{Khojiakbar Botirov$^{\mathrm{1}}$\orcidlink{0009-0004-6690-3365}, Hazem Sallouha$^{\mathrm{2}}$\orcidlink{0000-0002-1288-1023}, Sofie Pollin$^{\mathrm{2}}$\orcidlink{0000-0002-1470-2076} and Marcos Katz$^{\mathrm{1}}$\orcidlink{0000-0001-9901-5023}}
\IEEEauthorblockA{\textit{$^{\mathrm{1}}$Centre for Wireless Communications}, University of Oulu, 90570 Oulu, Finland\\ \textit{$^{\mathrm{2}}$Department of Electrical Engineering (ESAT) - WaveCoRE}, KU Leuven, 3000 Leuven, Belgium \\E-mails: \{khojiakbar.botirov, marcos.katz\}@oulu.fi, \{hazem.sallouha, sofie.pollin\}@esat.kuleuven.be}\\
\vspace{-3em}

\thanks{The present work has received funding from the European Union's Horizon 2020 Marie Skłodowska Curie Innovative Training Network Greenedge (GA. No. 953775). The work of  Hazem  Sallouha was funded by the  Research Foundation – Flanders (FWO), Postdoctoral Fellowship No. 12ZE222N.}
}


%


\maketitle


\begin{abstract}
In this work, we propose a Multiple Access Control (MAC) protocol for Light-based IoT
(LIoT) networks, where the gateway node orchestrates and schedules batteryless 
nodes’ duty-cycles based on their location and sleep time. The LIoT concept represents a sustainable solution for massive indoor IoT applications, offering an alternative
communication medium through Visible Light Communication
(VLC). While most existing scheduling algorithms for intermittent batteryless IoT aim to maximize data collection and enhance dataset size, our solution is tailored for environmental sensing
applications, such as temperature, humidity, and air quality
monitoring, optimizing measurement distribution and minimizing blind spots to achieve comprehensive and uniform environmental sensing. We propose a Balanced Space and Time-based Time Division
Multiple Access scheduling (BST-TDMA) algorithm, which addresses
environmental sensing challenges by balancing spatial and temporal factors to
improve the environmental sensing efficiency of batteryless LIoT nodes. Our measurement-based results show that BST-TDMA was able to efficiently schedule duty-cycles with given intervals.
\end{abstract}

\begin{IEEEkeywords}
LIoT, MAC, batteryless, IoT, BST-TDMA, Sustainability.
\end{IEEEkeywords}

\section{Introduction}
The Internet of Things (IoT) has revolutionized how we interact with technology, enabling seamless device connectivity. This transformation supports smart environments, automated systems, and data-driven decision-making across various domains, including healthcare, agriculture, industry, and smart cities \cite{s23167194}. As the number of connected devices grows exponentially, there is a pressing need for sustainable and efficient communication technologies to support large-scale IoT deployments.

Sensing technology aims to realize the "smart dust" concept, where billions of tiny, invisible computers enhance everyday life and infrastructure. Batteries represent a significant challenge to achieving a sustainable IoT due to their cost, size, environmental impact, and limited lifespan. Replacing and disposing of billions of batteries annually is both costly and environmentally irresponsible \cite{10.1145/3131672.3131699}.

Batteryless IoT nodes, powered by ambient light, align with the growing emphasis on environmentally friendly technologies and reduce maintenance costs associated with battery replacement. A batteryless Light-based IoT (LIoT) is a concept introduced by \cite{KATZ2019163176}, where indoor light can be both used to power nodes and leverages light as a communication medium, offering advantages such as energy efficiency, enhanced security, and reduced electromagnetic interference. These benefits make LIoT particularly suitable for indoor IoT applications where energy conservation and minimal interference are crucial. 
A LIoT node utilizes VLC to facilitate communication among IoT devices, and the usage of supercapacitors as an energy buffer in combination with organic photovoltaic (OPV) cells to harvest energy makes it a sustainable solution. Unlike traditional radio frequency (RF)-based networks, LIoT network can only use scheduled-based MAC such as Time Division Multiple Access (TDMA). Another challenge is LIoT nodes' intermittent operation nature, due to limited energy buffer. For batteryless IoT nodes, every duty-cycle should be used efficiently.
In response to these challenges, our work introduces the scheduling of LIoT networks. A robust and efficient scheduling mechanism is needed, which ensures reliable communication and maximizes the network's sustainable operation. We propose a novel Balanced Space and Time-based Time Division Multiple Access (BST-TDMA) scheduling algorithm, which optimizes duty-cycles in the LIoT network by considering the spatial distribution and duty-cycle of batteryless nodes.

The gateway node of an IoT network, composed of energy-limited nodes, orchestrates communication schedules based on node location and last duty-cycle time. This approach improves network efficiency, ensures nodes operate within energy constraints, and reduces communication blind spots, ensuring a uniform operation in this spatial and time domain.

The main contributions of the paper are:
\begin{itemize}
    \item Introducing a sustainable LIoT node energy consumption model based on node's power profiling.
    \item Developing a novel BST-TDMA scheduling algorithm optimized for environmental sensing. Comparing our solution to the conventional unbalanced scheduling.
    \item Reducing communication blind spots, ensuring comprehensive environmental monitoring. Statistical evaluation of the proposed BST-TDMA protocol.
\end{itemize}

The rest of this paper is organized as follows: Section~\ref{Sec_II} reviews related work. Section~\ref{Sec_III} describes the LIoT node model, indoor illumination model, and BST-TDMA. Section ~\ref{Sec_IV} evaluates the performance of BST-TDMA and Section~\ref{Sec_V} concludes with future work.

\section{Related work}
\label{Sec_II}
 Wireless sensing systems have revolutionized data acquisition and monitoring applications in many fields such as indoor environment sensing, smart offices, smart greenhouses, etc., Nonetheless, they usually experience a limited lifetime due to the expensive nature of wireless communication. Numerous works in the literature explore duty-cycle scheduling in energy-harvesting batteryless IoT nodes using RF as communication media.\\ In particular, several MAC protocols were proposed to extend network lifetime while optimizing throughput, latency, and fairness for radio media. 
 Wake-up radio-based MAC protocols such as IEEE 802.15.4, B-MAC, X-MAC\cite{10.1109/TNET.2014.2387314} have been proposed as better alternatives to always-on protocols due to their superior energy efficiency and channel utilization. In contrast to always-on protocols—in which sensor nodes continuously listen for or transmit data packets, they operate by systematically putting the node’s main radio into sleep mode, which is later woken up briefly to either receive or transmit data. Wake-up radio-based MAC can be done in a synchronous or asynchronous manner. Synchronous duty-cycled MACs keep a common time reference among the nodes, which introduces time synchronization overhead and complexity. On the other hand, the asynchronous counterparts utilize schemes like preamble sampling, random duty-cycling, or receiver initiation to circumvent synchronization challenges \cite{ricardo}. Notwithstanding, asynchronous MACs are still susceptible to idle listening energy consumed listening for data packets during active periods without success, and other major issues like time spent waiting for sleeping nodes to wake up, and overhearing energy consumed receiving data meant for another node \cite{10.1109/TNET.2014.2387314}. 
 Advances in wake-up radio technology have provided ways to resolve most of the challenges faced by duty-cycled MAC protocols. Wake-up radios are ultra-low-power (ULP) receivers with orders of magnitude lower power consumption compared to existing low-power radio transceivers. However, for a batteryless LIoT node, such a solution is not applicable due to the limited energy buffer. 
 Authors of \cite{10.1145/3649221} present Greentooth, a robust and energy-efficient wireless communication protocol for intermittently powered LIoT nodes. It enables reliable communication between a receiver and multiple batteryless sensors using TDMA-based scheduling and low-power wake-up radios for synchronization. 
MAC protocols for intermittent operating sensor networks mainly focus on radio communication-based applications and try to maximize the number of duty-cycles, whereas we propose MAC protocol using concepts of duty-cycled scheduling via utilization of VLC. Our custom data exchange protocol is an alternative solution to signify the uniform distribution of the environmental sensing measurements. 

\section{System Design}
\label{Sec_III}
In this section, we present a quick overview of the batteryless LIoT node model under consideration, as well as how the indoor illumination was modeled. Also, we will describe the simulation of the environment and present the BST-TDMA MAC protocol.

\subsection{Batteryless LIoT model}
\begin{figure}[htbp]
\vspace{-0.5em}
\centering
\includegraphics[scale=0.35]{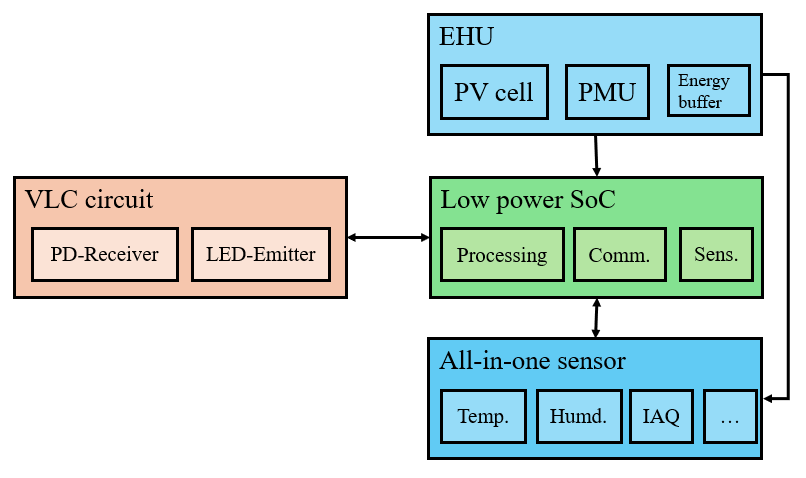}
\vspace{-0.5em}
\caption{Principal components of a batteryless LIoT sensor node.}
\vspace{-1em}
\label{fig:net_diagram}
\end{figure}

The block diagram in Fig. \ref{fig:net_diagram} presents the principal components of a batteryless LIoT node. The blocks illustrate the Energy Harvesting Unit (EHU), the all-in-one low-power sensor, the low-power System-on-Chip (SoC), and the integrated VLC circuit. Next, we present a description of the main characteristics of each of these components, focusing on enhancing the reliability of the batteryless LIoT sensor nodes.

The EHU plays a crucial role in supplying energy by utilizing organic photovoltaic (PV) cells and powering electronic circuits for sustainable energy harvesting. The Power Management Unit (PMU) within the EHU follows a Harvest-Storage-Use (HSU) strategy, which allows the system to activate immediately after gathering enough energy. This is achieved by optimizing energy transfer through Maximum Power Point Tracking (MPPT) to a supercapacitor (SCap) that serves as an energy buffer. SCaps are preferred over batteries due to their almost infinite charge cycles, high charge-discharge efficiency (97\%-98\%), high power density, and lack of heat generation during discharge. In our implementation, we used Epishine LEH OPV in the EHU \cite{epishine}. Low-power sensors, such as the Bosch BME680 \cite{bme680} integrated into our LIoT node, are essential for monitoring indoor air quality (IAQ), measuring factors like temperature, humidity, air pressure, gases, and CO2 while complying with the ISO16000-29 standard for VOC detectors, all with minimal energy consumption \cite{landivar2024batterylessblelightbasediot}. These sensors must align with the latest ultra-low-power IoT solutions. The low-power SoC integrates components like memory and both analog and digital peripherals into a single substrate, managed by a microprocessor. ARM Cortex M series-based SoCs are preferred for their low-power design, making them ideal for applications with strict power constraints. BLE is often chosen for low-power, short-range communication, while alternative methods like VLC complement RF communications, enhancing the reliability of batteryless IoT nodes by covering the light spectrum from 380 nm to 780 nm and improving the overall reliability of indoor short-range networks \cite{10283657}. The determination of sleep and active cycles in the proposed solution is based on the energy consumption analysis of the LIoT nodes.
This analysis employs the producer-consumer model for autonomous wireless devices presented in~\cite{Martinez2015}. The model considers all the energy sources and energy consumed by a wireless IoT node, and based on the energy conservation principle, outlines that the energy produced in a total operational time $T$ should be equal to or greater than the energy consumed, which is described in the following inequality: 
\\
\begin{equation}
E_{buf}\!+\!\underbrace{\int_{0}^{T}\!P_{harv}(\tau)d\tau}_{E_{harv}(T)}\geq\underbrace{\int_{0}^{T}\!P_{dev}(\tau)d\tau}_{E_{dev}(T)}
\label{eq1}
\end{equation}
\\
where $\tau$ is used as the variable of time, $E_{buf}$ represents the initial energy in the buffer in Joules, $P_{harv}(\tau)$ is the harvested power for an instant of time $\tau$ in Watts, and $P_{dev}(\tau)$ is the power consumed by the device for an instant of time $\tau$ in Watts.
From Eq.(\ref{eq1}) we can calculate the necessary sleep time \(T_s\) period to maintain continuous operation of the node. By using the whitepaper of EHU manufacturer \cite{epishine}, we can calculate the EHU harvested power \(P_{harv}\) based on the linear relationship between the \(P_{harv}\) and illuminance. Using the measurements of the LIoT node power profiling, we measured \(E_{dev}\) = 0.2233 J during duty-cycle lasting dur = 4.45 seconds. Also, our LIoT nodes consume \(P_{sleep}\) = 0.28mW power during sleep time. Based on our measurements, we can calculate \(T_s\) as Eq.(\ref{eq: t_sleep}). 
\\
\begin{equation}
     T_s = E_{dev}/(P_{harv} - P_{sleep})
\label{eq: t_sleep}
\end{equation}
\\
In each duty-cycle of LIoT nodes, data is exchanged with the gateway node by using infrared LEDs for uplink, and photodiodes for downlink. LIoT senses and sends environmental observations, including illumination, temperature, humidity, IAQ, and air pressure. When the gateway node orchestrates the next duty-cycle of the LIoT node by directly calculating \(T_s\) and sending it to the LIoT node without taking into account measurements from other nodes, this approach is called Unbalanced STDMA (U-STDMA). More details about the LIoT design may be found in our earlier work \cite{landivar2024batterylessblelightbasediot}.
\subsection{Indoor illumination modelling}
\begin{figure}[htbp]
\vspace{-0.5em}
\centering
\includegraphics[scale=0.29]{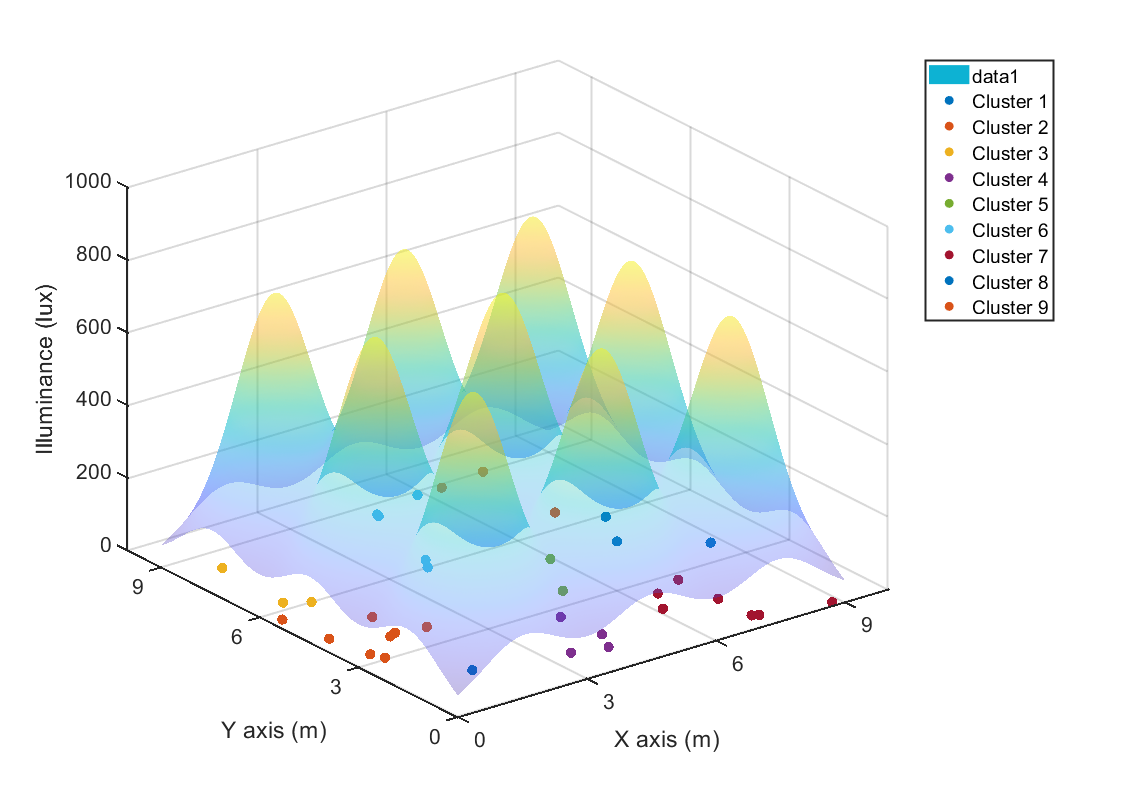}

\caption{Clustered LIoT node distribution and Illuminance colormap.}
\vspace{-1em}
\label{fig:node_dist}
\end{figure}

To develop and evaluate the proposed BST-TDMA with a larger number of LIoT nodes we modeled indoor illumination. Simulating the illumination, in a 3x3x3 meter room with nine ceiling-mounted light points functioning as Access Points (AP) involves analyzing the distribution of light to ensure optimal coverage and uniformity. Each light point is strategically placed on a 3x3 grid pattern on the ceiling to maximize the area of illumination and serve dual purposes: providing light and facilitating communication as APs for a LIoT network. The simulation considers factors such as the luminous intensity, beam angle, and Lambertial radiation index of the light source to evaluate how effectively the room is illuminated. The goal is to achieve a focused irradiance of the APs to segment the room surface into space clusters. Space clusters are represented by each AP, which is located in its centroid point and predefined boundaries were given to allocate LIoT nodes into a particular cluster. We assume a set of LIoT nodes in each cluster observe the same environmental phenomenon at a given instant of time. This assumption leads us to do space-based scheduling of LIoT nodes.
Fig. \ref{fig:node_dist} represents the illumination map, where the location of every AP can be seen. Also, the distribution of random 36 LIoT nodes at the same height was clustered to one of the nine space clusters.
\begin{table}[htbp]
    \centering
    \begin{tabular}{@{}llc@{}}
        \toprule
        \textbf{Variable} & \textbf{Description} & \textbf{Value} \\ 
        \midrule
        \(W_t\) & Width (meters) & 9 \\ 
        \(H_t\) & Height (meters) & 3 \\ 
        \(L_t\) & Length (meters) & 9 \\ 
        \(N_{\text{AP}}\) & Number of Access Points & 9 \\ 
        \(\text{pos}_{\text{led}}\) & Access Point locations & \(\begin{bmatrix}1.5, 1.5, 3; \\ 1.5, 4.5, 3; \\ 1.5, 7.5, 3; \\ 4.5, 1.5, 3; \\ 4.5, 4.5, 3; \\ 4.5, 7.5, 3; \\ 7.5, 1.5, 3; \\ 7.5, 4.5, 3; \\ 7.5, 7.5, 3\end{bmatrix}\) \\ 
        \(I_v\) & Luminous flux (lumen) & 3200 \\ 
        \(\phi_{0.5}\) & Radiation semi-angle (degrees) & 30 \\  
        \bottomrule
        \\
    \end{tabular}
    \caption{Parameters of the indoor illumination modeling}
    \label{tab:variables}
\end{table}

From Table \ref{tab:variables}, we can notice that the radiation semi-angle was set at \( 30^\circ \) and Eq.(\ref{eq: lambertian}) represents Lambertial radiation index to achieve homogeneous illuminance at the surface of each cluster.
\\
\begin{equation} 
m = -\frac{\ln(2)}{\ln\left(\cos\left(\frac{\phi_{0.5} \cdot \pi}{180}\right)\right)}.
\label{eq: lambertian}
\end{equation}
\\
The radiation vector (also known as the light vector) indicates the direction from which light is coming and is defined as \(R = [0,0, -1]\).
\begingroup
\small
\begin{equation}
D = [x_{\text{AP}} - x_{\text{LIoT}}, y_{\text{AP}} - y_{\text{LIoT}} ,z_{\text{AP}} - z_{\text{LIoT}}]
\label{eq: d_AP}
\end{equation}
\endgroup
The distance vector \(D\) between each AP and LIoT was calculated by Eq.(\ref{eq: d_AP}) and we can obtain the distance between AP and LIoT by \(d_{\text{AP}} = ||D||\).
\begin{equation}
\phi = \arccos\left(\frac{\mathbf{R} \cdot \mathbf{D}}{\|\mathbf{R}\| \|\mathbf{D}\|}\right)
\label{eq: phi}
\end{equation}
The angle of incidence of the light \(\phi\) on the LIoT node was obtained by using the dot product of \(R\), \(D\), and the magnitudes of these vectors, Eq. (\ref{eq: phi}).
\begin{equation}
    \text{I}_{\text{AP}} = I_v \left(\frac{m+1}{2\pi d_{\text{AP}}^2}\right) \cdot \left|\cos(\phi)^m\right|
\label{eq: illum}
\end{equation}
The illuminance on the surface of the LIoT from a single AP was calculated by Eq.(\ref{eq: illum}). The total illuminance at the position of particular LIoT nodes was obtained by \(I_{\text{total}} = \sum_{i=1}^{N_{\text{AP}}} I_{\text{AP},i}. \)

\subsection{BST-TDMA algorithm}
This algorithm performs LIoT node scheduling in a defined space. A scheduling vector is generated with time increments, and nodes are scheduled in a manner that avoids duty-cycle conflicts. The scheduling algorithm iterates through possible time slots, ensuring no overlapping events occur within the specified end time. The sleep time \(T_s\) of the LIoT was calculated respecting Eq.(\ref{eq: t_sleep}). Each space cluster contains a set of LIoT nodes represented by matrix \(node_{C}\), where the 1st and 2nd rows indicate LIoT ID and its \(T_s\). The ideal spatial distribution of the set of LIoT nodes with the maximum amount of duty-cycles for a given period would be, if they are located in the same position and have \(T_s\) equal to the minimum. By the above-mentioned case, we calculate the ideal periodicity \(per = min(T_s)/length ({node_{C}}(1))\) of duty-cycles, which maximizes the number of measurements and tries to achieve a uniform distribution of the measurements. 

\begin{algorithm}
\caption{BST-TDMA}\label{alg:cap}
\begin{algorithmic}
\WHILE{True}
    \STATE updated $\gets$ false
    \FOR{$i = 1$ to length\((node_{C})\)}
        \STATE \(event_{id}\) $\gets$ node$_{id}(i)$
        \STATE \(last_{time}\) $\gets$ result(2, result(1,:) == \(event_{id}\))
        \STATE \(next_{time}\) $\gets$ \(last_{time}\)(end) + node$_{C}(i)$ + \(dur\)
        \WHILE{\(next_{time}\) $\leq$ \(end_{time}\)}
            \IF{not has\_conflict(result, \(next_{time}\), \(per\))}
                \STATE result $\gets$ [result, [\(event_{id}\); \(next_{time}\)]]
                \STATE updated $\gets$ true
                \STATE \textbf{break}
            \ELSE
                \STATE \(conflicting_{time}\) $\gets$ result(2, abs(result(2,:) -\(next_{time}\)) $<$ \(per\))
                \STATE \(d\) $\gets$ $\min$($\text{abs}$(\(conflicting_{time}\) - \(next_{time}\)))
                \STATE \(next_{time}\) $\gets$ \(next_{time}\) + (\(per\) - \(d\))
            \ENDIF
        \ENDWHILE
    \ENDFOR
    \IF{not updated}
        \STATE \textbf{break}
    \ENDIF
\ENDWHILE
\end{algorithmic}
\end{algorithm}
From Alg.\ref{alg:cap}, we can see that BST-TDMA allocates the next duty-cycle to a LIoT node by adding \(T_s\) + dur (duty-cycle duration) and checks if the next\_time has a conflict with other nodes' duty-cycles. When the conflict is detected, the algorithm calculates the time variable d, which works as an additional period to ensure a delay between two duty-cycles equal to \(per\).
\begin{algorithm}
\caption{Check for Conflict}\label{alg: conf}
\begin{algorithmic}
\item[] has\_conflict(schedule, event\_time, \(per\)):
\FOR{i = 1 to $length(schedule)$}
    \STATE \(scheduled_{time}\) $\gets$ schedule(2, i)
    \IF{abs(\(scheduled_{time}\) - event\_time) $<$ per}
        \STATE conflict $\gets$ true
        \STATE \textbf{return} conflict
    \ENDIF
\ENDFOR
\STATE \textbf{return} conflict
\end{algorithmic}
\end{algorithm}
The BST-TDMA uses the has\_conflict function to check if the next scheduled duty-cycle is reserved in a time slot exceeding \(per\) Alg.\ref{alg: conf}.

\section{Performance Evaluation}
\label{Sec_IV}
\begin{figure}[htbp]
\vspace{-0.49em}
\centering
\includegraphics[scale=0.49]{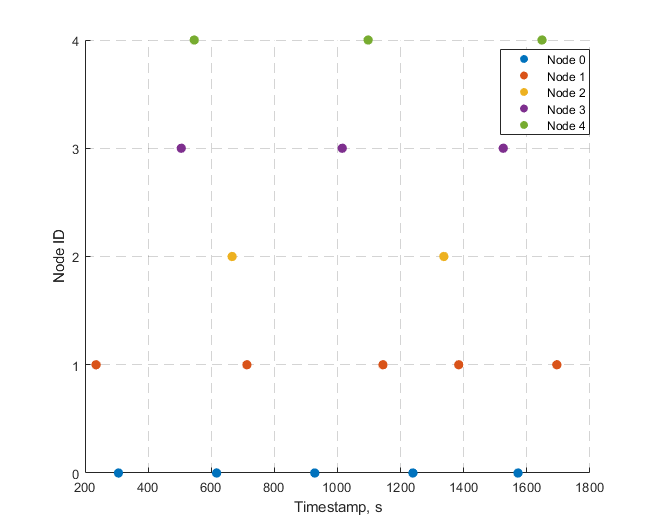}
\vspace{-0.5em}
\caption{LIoT nodes scheduling of a single cluster.}
\vspace{-1em}
\label{fig:duty_cycles}
\end{figure}

In this section, we present the performance evaluation of the BST-TDMA method. We compare it with the unbalanced scheduling approach presented in \cite{landivar2024batterylessblelightbasediot}, which maximizes the number of duty-cycles without knowing measurement distribution over space-time. We assume that the duty-cycles of a set of LIoT nodes per cluster are scheduled by the gateway node using Alg.\ref{alg:cap}. A visualization of the LIoT nodes' transmissions over time when using BST-TDMA is presented in Fig. \ref{fig:duty_cycles}, in which one can see the set \(node_{C}\) of 5 LIoT nodes with \(T_s=[306,235,666,505,546]\) and \(per=47\) can be seen. Fig.\ref{fig:duty_cycles} illustrates every LIoT node schedule for a duration of 30mins and intervals between duty-cycles of the same node are not equal.

\begin{figure}[htbp]
\vspace{+0.1em}
\centering
\includegraphics[scale=0.27]{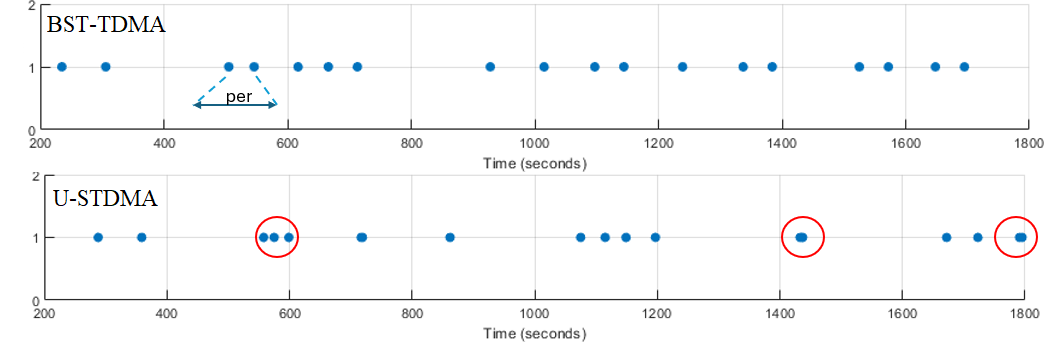}
\vspace{-0.5em}
\caption{Combined LIoT schedule.}
\vspace{-1em}
\label{fig:schedule_compare}
\end{figure}

The difference between unbalanced and Balanced STDMA is the variable \(per\), which ensures the exclusion of the same measurements in the time domain. Uniformly distributed measurements provide a higher quality of environment sensing. Fig.\ref{fig:schedule_compare} depicts the combined LIoT schedules for the unbalanced and proposed balanced approach, noting the larger amount of blind durations and duty-cycle intervals being minimum between LIoT nodes in U-STDMA. BST-TDMA, on the other hand, guarantees that the minimum periodicity between duty-cycles is equal to \(per\). 

\begin{figure}[htbp]
\vspace{-0.5em}
\centering
\includegraphics[scale=0.46]{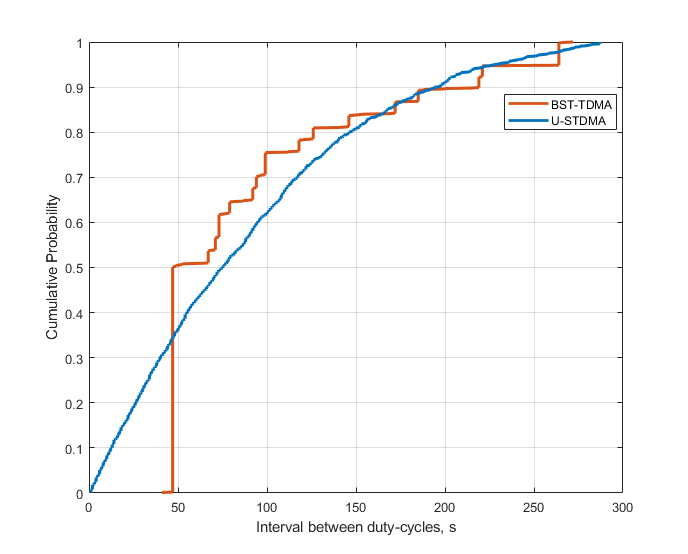}
\vspace{-0.5em}
\caption{BST-STDMA vs U-STDMA CDF}
\vspace{-1em}
\label{fig:cdf}
\end{figure}

Depiction of the Cumulative Distribution Function (CDF) of the duty-cycle intervals of both scheduling algorithms gives us insights to understand better the effect of the considered solution. From Fig.\ref{fig:cdf}, it can be observed, that BST-TDMA scheduled at least 50\(\%\) of the duty-cycles at periodicity equal to \(per\) for the same set of LIoT \(node_{C}\) with a duration of the measurement equal to 24 hours. From the unbalanced scheduling CDF, we can observe the randomness of the duty-cycle distribution, which significantly affects the practicality of this approach. 

\begin{figure}[htbp]
\vspace{-0.5em}
\centering
\includegraphics[scale=0.49]{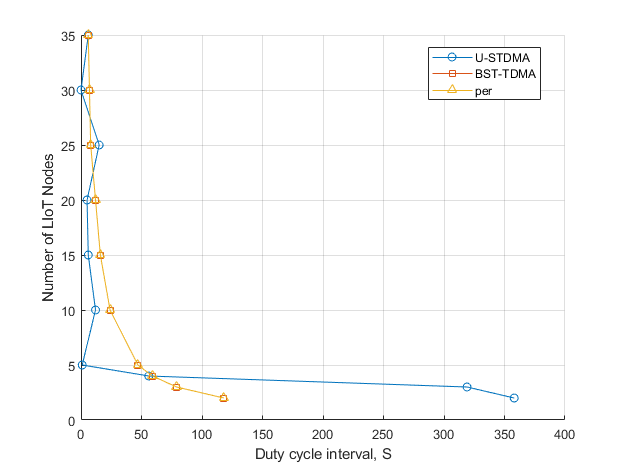}
\vspace{-0.5em}
\caption{duty-cycle interval mode comparison by number of nodes.}
\vspace{-1em}
\label{fig:stats}
\end{figure}

\begin{figure}[htbp]
\vspace{-0.5em}
\centering
\includegraphics[scale=0.5]{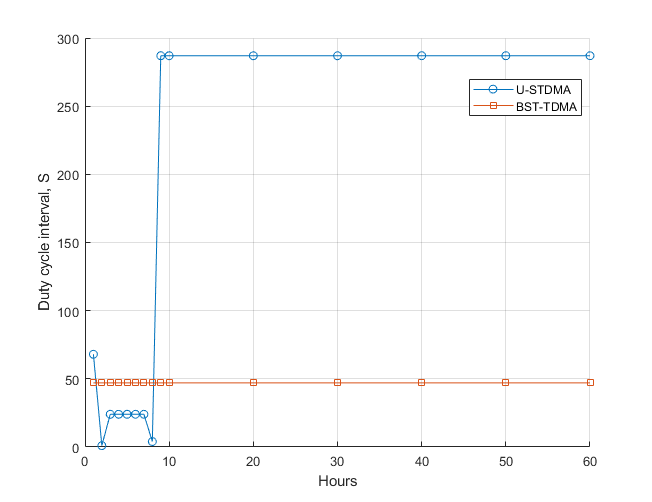}
\vspace{-0.5em}
\caption{duty-cycle interval mode comparison by duration.}
\vspace{-1em}
\label{fig:hours_mode}
\end{figure}

A comparison of both scheduling approaches with a large number of LIoT nodes shows, that the mode of the scheduled duty-cycle intervals for BST-TDMA is always equal to calculated \(per\), while the U-STDMA shows randomness in this term, Fig.\ref{fig:stats}. Another significant difference between the two MAC protocols is found while testing them with increasing measurement time from 1 to 60 hours. Fig.\ref{fig:hours_mode} shows, that the mode interval between duty-cycles is always equal to \(per\) in the case of BST-TDMA, and for the U-STDMA, it increases more than 5 times over \(per\) after 9 hours for the same set of LIoT \(node_{C}\).

\section{Conclusion}
\label{Sec_V}
In this paper, we introduced the BST-TDMA duty-cycle scheduling algorithm for intermittent operating LIoT nodes, which optimizes the duty-cycle schedule taking into account the spatial distribution of LIoT nodes to enhance network efficiency, reduce communication blind spots, and ensure comprehensive environmental sensing.
The simulation results demonstrate the effectiveness of the BST-TDMA algorithm compared to the conventional Unbalanced scheduling approach. Our approach offers a sustainable and scalable solution for large-scale indoor IoT applications, addressing the challenges associated with batteryless LIoT nodes and VLC-based communication networks.
Future work will focus on further refining the BST-TDMA algorithm to adapt to dynamic environmental conditions and varying node densities. Additionally, we plan to explore the duty-cycle scheduling based on the type of environment observation.
\bibliographystyle{ieeetr}
\bibliography{bibliography}

\end{document}